\documentclass[]{aa501}
\usepackage{graphicx}

\begin{document}

\title{Magnetic fields and radio polarization of barred galaxies}  
\subtitle{3D dynamo simulations}
\author{K. Otmianowska-Mazur\inst{1}
\and D. Elstner\inst{2}
\and M. Soida\inst{1}
\and M. Urbanik\inst{1}
}
\institute{Astronomical Observatory, Jagiellonian
University, Krak\'ow, Poland
\and Astrophisikalisches Institut Postdam, Germany
}
\offprints{K. Otmianowska-Mazur}
\mail{otmian@oa.uj.edu.pl}
\date{Received 28 August 2001; Accepted 5 December 2001}

\abstract{ A three-dimensional (3D) MHD model is applied to simulate the
evolution of a large-scale magnetic field in a barred galaxy possessing
a gaseous halo extending to about 2.8~kpc above the galactic plane.
As the model input we
use a time-dependent velocity field of molecular gas resulting from
self-consistent 3D N-body simulations of a galactic disk. We assume that the
gaseous halo rotates differentially co-rotating with the disk 
or decreasing its velocity in the Z direction.
The dynamo process included in the model yields the
amplification of the magnetic field as well as the formation of field
structures high above the galactic disk. The simulated magnetic fields
are used to construct the models of a high-frequency (Faraday rotation-free)
polarized radio emission that accounts for effects of projection
and limited resolution, and is thus
suitable for direct comparison with observations.\\ We found that the resultant
magnetic field correctly reproduces the observed structures of polarization
B-vectors, forming coherent patterns well aligned with spiral arms and with the
bar.  The process initializing a wave-like behavior of the magnetic field,
which efficiently forms magnetic maxima between the spiral arms,
is demonstrated. The inclusion of the galactic halo constitutes a step
towards a realistic model of galactic magnetic fields that includes
as many dynamical components as needed for a realistic description.  
}
\authorrunning{Otmianowska-Mazur}
\titlerunning{Magnetic fields of barred galaxies}
\maketitle

\section{Introduction}

Barred galaxies offer a unique opportunity to study the interrelations between
gas flows and magnetic field structure.  The bars are known to excite strong
spiral patterns in the interstellar gas  and to cause strongly
non-axisymmetric gas flows (Englmaier \& Gerhard \cite{engl97}).  
Recent observations show that the magnetic field is strongly
modified by gas shearing motions, although the field perturbations do not
coincide with the suspected position of gas velocity perturbations,
as expected from simple MHD models.
Some concepts suggest that the non-axisymmetric gas flows in the bar
may constitute the main mechanism amplifying the galaxy-scale magnetic fields
(Lesch \& Chiba \cite{les97}). Studying strong gaseous spiral 
patterns driven by the bar perturbation
also offer a unique opportunity to understand the relationships between
spiral arms and magnetic fields that are currently lively debated.

The question of the gas reaction to the bar potential has a long history
(Moss et al. \cite{mosetal98}, Moss et al. \cite{mos99} and
references therein).  However, little has
been done to study the evolution of magnetic field in such objects. Moss
(\cite{mos98})
analyzed the magnetic field evolution in barred galaxies, however his models
are restricted to two dimensions, while the vertical
structures are essential for the magnetic field evolution.
Regular magnetic fields are not only passively deformed by large-scale gas
motions in the galactic plane but their evolution is also driven by turbulent
motions via the dynamo mechanism. The latter process generates
galaxy-scale poloidal (hence also vertical) magnetic field components.
Though this genuinely three-dimensional magnetic field structure
can still be analyzed in two dimensions by applying a solenoidality
condition, we find it of great interest to apply a fully 3D model of
the field evolution. Moreover we
also get an essential vertical component of the large-scale gas motion,
which requires a completely three-dimensional treatment.
A three-dimensional analysis of the interaction of the dynamo
process with spiral
arms has been performed by Rohde \& Elstner (\cite{roh98}) and Rohde et al.
(\cite{roh99}).  However, they considered only
the turbulence enhancement in star-forming parts of spiral arms with no
compression effects included, which is appropriate for the weak density wave
objects (NGC 6946, Beck \& Hoernes~\cite{bec96}) they intended to explain.
A comprehensive study of
interactions of magnetic fields with bar-induced gas flows and spiral arms
involving the dynamo process (hence carried out in three dimensions) is still
urgently needed.

In our previous work (Elstner et al. (\cite{els00}),
v. Linden et al. (\cite{lin98}), Elstner et al. (\cite{els98})) 
we analyzed the magnetic field evolution under the
influence of the bar and spiral arms.  However, these studies assumed a thin
gaseous disk only. A large gaseous halo may be of crucial importance
for the problem.  The dynamo process includes the generation of large-scale
poloidal fields then transformed into azimuthal ones by differential
rotation.  A realistic analysis of the magnetic
field evolution under a combined effect of the gas flows and the turbulent
dynamo needs to involve an extended gaseous halo.

For these reasons we performed the study of magnetic field evolution assuming
a galaxy composed of a disk and of a rarefied ionized halo
with a scale height of 2.8~kpc. This expands our previous results involving 
a thin disk only. As previously, we use a realistic model of the gas 
flow, by analyzing motions of inelastically-colliding gas clouds in the
self-consistent calculated gravitational potential of stars and gas.
We adopted conditions suitable for the formation of a stellar bar. We
assumed the magnetic field to be partially coupled to the gas via the turbulent
diffusion process. We checked the results for all the assumptions by varying 
the diffusion coefficient, the strength of the dynamo action, and also
by switching off the turbulent dynamo.

The magnetic field in external galaxies is usually studied by means of the
radio polarization, yielding the view of the magnetic field integrated along
the line of sight and convolved to a certain radio telescope beam. As our
models yield complex three-dimensional magnetic field structures, we decided to
analyze our magnetic field structures by simulating the polarization maps,
which yields much more concise information, easy to compare with observations.

\noindent
\section{The model}

\subsection{Numerical methods}

Our experiment consists of two basic steps: the three-dimensional N-body,
sticky particle calculations yielding a realistic gas velocity field and a
magnetohydrodynamical (MHD) model of the magnetic field evolution using the gas
flow picture from the first simulations 
(Elstner et al. \cite{els00}, v. Linden et al. \cite{lin98},
Otmianowska-Mazur et al. \cite{otm97}).  
The model galaxy consists of two essential components:
a collision-less self-gravitating disk composed of star clusters
as well as highly inelastic colliding gas clouds moving in
the gravitational potential of stars and gas, as well as an analytical
contribution due to the bulge and halo.
The code involves a sticky-particle cloud scheme as described by
Combes \& Gu\'erin (\cite{com85}). In our model the clouds were assumed to 
represent the large-scale velocity field of the gas
(see Otmianowska-Mazur et al. \cite{otm97}
for the detailed model description). We modeled the magnetic field evolution by
analyzing numerically the time-dependent solutions of the dynamo equation:

 $${\partial\vec{B}/\partial t=\hbox{rot}(\vec{v}\times\vec{B})+\hbox{rot}
(\alpha~ \vec{B}) -\hbox{rot}(\eta~\hbox{rot}\vec{B})}$$ 
\noindent
where $\vec{B}$ is the magnetic induction, $\vec{v}$ is the large-scale 
velocity
of the gas, $\alpha$ is the dynamo coefficient (a tensor in general,
having a diagonal form in our case) describing the mean helicity of
turbulent motions and $\eta$ is the coefficient of a
turbulent diffusion.

Calculations of the magnetic field are done with a modified version of
the ZEUS3D code (Stone \& Norman, \cite{sto92}) neglecting the
action of the magnetic field on the mean gas flow.
The effects of the magnetic
diffusion as well as the term of the $\alpha$-effect are introduced
directly into the code (von Linden et al.~\cite{lin98},
Elstner et al. \cite{els00}). 
We introduced the galactic halo of ionized gas extending to 2.8~kpc
above the galactic plane. This value was chosen due to computational constraints
and is four times larger than the disk thickness used in 
our earlier works (von Linden et al.~\cite{lin98},
Elstner et al. \cite{els00}).

To analyze the general properties of the magnetic field we decided to use the
simulated maps of polarized intensity, constructed on the basis of our results.
We performed it in the way similar to that described in our previous works (e.g.
Otmianowska-Mazur et al. \cite{otm00}) 
namely, by integrating the Stokes parameters I, Q and U along the line of sight,
convolving them with an assumed beam and computing the maps of polarized
intensity and the angle of polarization B-vectors.
This process yields information averaged over the line of sight, hence more
concise, and directly comparable with observations.

\subsection {Model assumptions and input parameters}

The clouds span the mass range between $5\,\times\,10^2$ and
$5\,\times\,10^5\,{\rm M}_\odot$ distributed in 10 mass-bins with logarithmic
intervals.  The disk of the galaxy consists of 38\,000 star clusters
(about $2\cdot 10^6$\,M$_\odot$ each)
and 38\,096 clouds.
The time step of calculations is $10^7$ years.  The model parameters adopted in
the computations are summarized in Table~\ref{tab1}.

\begin{table}
\caption[]{\label{tab1}Input parameters for N-body model}
\begin{flushleft}
\begin{tabular}{llllllllllllll}
\hline\noalign{\smallskip}
Model    &   \\
\hline\noalign{\smallskip}
Mass in $10^{10} {\rm M}_\odot$:     &\\
- dark-halo $M_h$               & 9.6\\
- disk $M_d$                 & 7.2 \\
- gas $M_g$                    & 1.6\\
- bulge $M_b$                 & 4.8\\
- $(M_d+M_g)/M_{\rm tot}$ & 0.38\\
Scale length in kpc: &\\
- dark halo       &15\\
- bulge           &1.2\\
- disk            &6\\
\noalign{\smallskip}
\hline
\end{tabular}
\end{flushleft}
\end{table}

\begin{figure*}
\resizebox{\hsize}{!}{\includegraphics{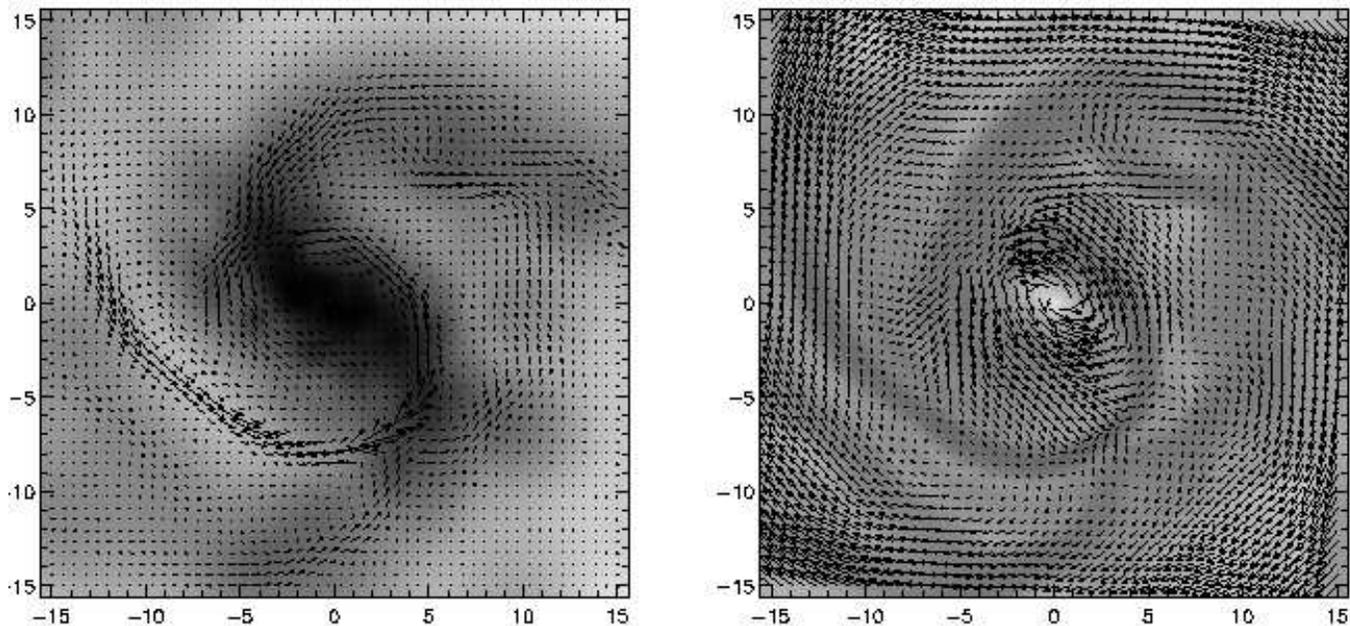}}
\caption{Magnetic field and gas density in the midplane for model A1 at
0.35~Gyr (left). Velocity vectors in the corotating frame with the bar 
and magnetic energy in greyscale at the midplane (right).
}
\label{bfeld}
\end{figure*}

\begin{table*}
\caption[]{\label{tab2}Input parameters for MHD models}
\begin{flushleft}
\begin{tabular}{llllllllllllllllllllllllllllllllllll}
\hline\noalign{\smallskip}
Model    & A1 & A2 & A3 & B  & C  \\
\hline\noalign{\smallskip}
$\alpha_{0xx}=\alpha_{0yy}$ [km/s]: & 10 & 1.0 & 0.1 & 0  & 10\\
$\alpha_{0zz}$ [km/s]: & $-$10 & $-$1.0 &  $-$0.1&  0  & $-$10 \\
$\eta_0$ $ [cm^2/s$]: & $3\times 10^{26}$ & $3\times 10^{25}$ & $3\times
10^{25}$ & $3 \times 10^{26}$  & $3 \times 10^{26}$ \\
rotation of halo in Z direction: &
const&const&const&const& decreasing\\
\noalign{\smallskip} \hline
\end{tabular}
\end{flushleft}
\end{table*}

The kinematic evolution of the magnetic field is solved in a 3D rectangular
grid of points, where the XY plane is the galactic plane and the Z axis
is the axis of rotation.  For all experiments we use 0.3~kpc as a grid
step along the X and Y axes and 0.08~kpc along the Z axis.
The disk scale height is 0.8~kpc, and halo height is 2.8~kpc above
the galactic plane. The number of grid points in X and Y directions
is 101, while in Z, 71 points. The adopted time step is $0.5\times10^5$~yr. 

We performed simulations of magnetic field evolution in a barred gas-rich
galaxy with five different sets of parameters. The input parameters for all
models are summarized in Table \ref{tab2}. We include the diagonal
components of the $\alpha$ tensor changing sinusoidally in the disk
along the Z-direction with the maximum
value of $\alpha_0$ presented in Table \ref{tab2}. In the halo we assume
$\alpha_{xx}=\alpha_{yy}=\alpha_{zz}=0 $ above 1~kpc.
In order to saturate the dynamo process  at equipartition field strength
$B_{eq}$ we adopt the usual $\alpha$-quenching
with $\alpha = \alpha_0/(1+(B/B_{eq})^2)$ (Elstner et al. 2000).
The diffusion coefficient $\eta$ grows linearly in the Z-direction,
which is the result of the hydrostatic equilibrium assumption on the galactic
disk (Fr\"ohlich \& Schultz, \cite{fro96}). Its midplane value  of
$3 \times 10^{26} \rm cm^2/s$ results from a turbulent velocity of 
10~km/s and a correlation time of $3 \times 10^7 \rm years$. 
Due to uncertainty in the correlation time and its possible reduction by 
the magnetic fields we assume that the value of $\eta$ is ten times
smaller in experiments A2 and A3 (without $\eta$-quenching).
In A1--B models we apply differential rotation which is independent
of the height above the disk plane up to 2.8~kpc. Such an assumption
is based on the observational results derived by different authors
that the Galactic halo co-rotates or nearly co-rotates with the disk
up to 3--4~kpc above the galactic disk (Kalberla \& Kerp \cite{kal98} and
references therein). In model C the speed of the differential rotation in the
halo decreases exponentially with increasing distance from the disk boundary
to half of its disk value, according to recent observation of NGC~5775
(T\"ullmann et al. \cite{tul00}).
In order to study possible changes of the magnetic field structure caused
by the bar and spiral arms we start our simulations from an already-evolved
magnetic field configuration computed using the initial velocity field.

\begin{figure*} 
\resizebox{\hsize}{!}{\includegraphics{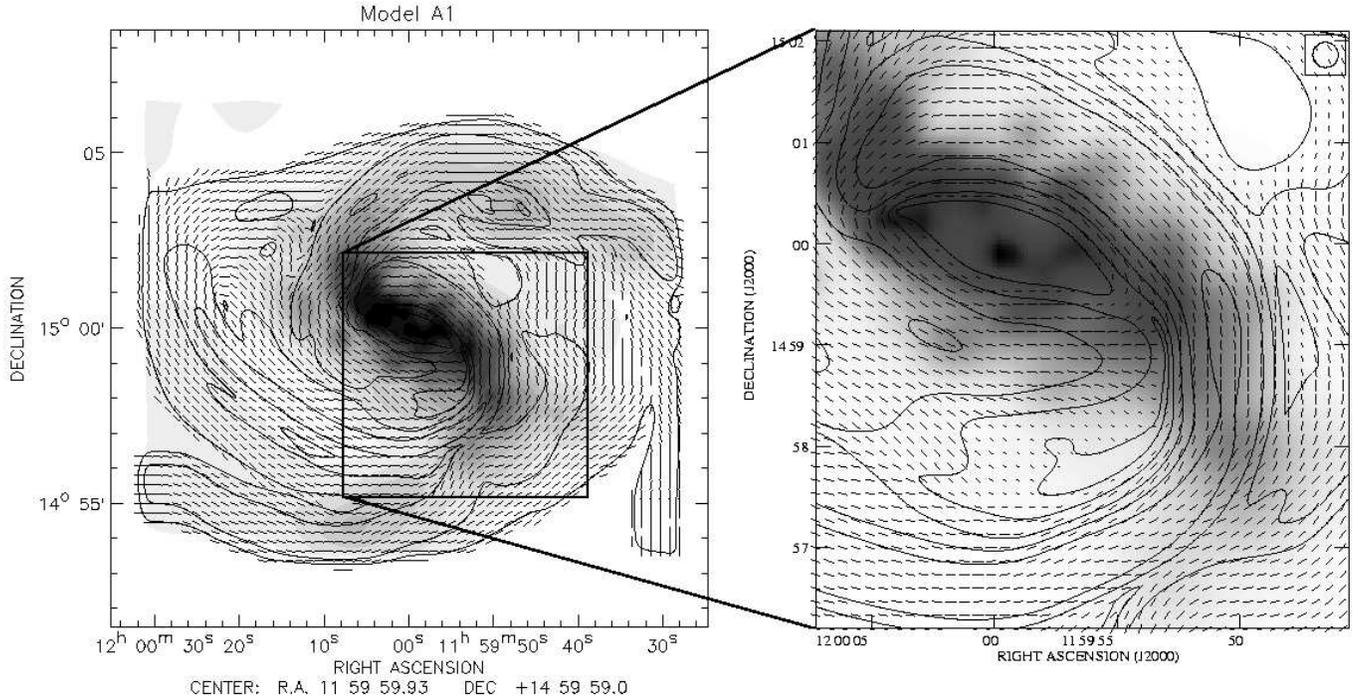}} 
\caption{Polarization map of model A1 at time 0.35G~yr and inclination
of $30^\circ$ superimposed onto gaseous map
(grey plot) integrated along the line of sight. Enlargement of the
most interesting region is shown on the right.}
\label{fmaps35} 
\end{figure*}

The polarization maps at selected time steps of our models are constructed with
the inclination of $30^\circ$ (almost face-on) and $85^\circ$ (almost edge-on).
The model assumes the relativistic electrons with an energy spectral index
$\gamma = 2.8$. Their distribution is assumed to change vertically
as a Gaussian with the scale-height of 1~kpc. 
Radially, the electron density is uniform up to $R=12$~kpc.
We adopt 70\% as a
maximum polarization degree of synchrotron radiation (Pacholczyk \cite{pac70},
Otmianowska-Mazur et al. \cite{otm00}) and neglect any Faraday and
absorption effects. The maps are convolved with a beam of 10$\arcsec$
and 15$\arcsec$ for the edge-on (85$\degr$) and face-on (30$\degr$)
inclination, respectively.

\section{Results}

\subsection{Magnetic field structure in selected planes}

The bar turns rigidly with a corotation radius of about 8~kpc.
The gaseous spiral arms are excited by a bar perturbation but
they follow the bar at a somewhat lower angular speed.
They break into parts, some of them again join the opposite bar end while the
others slowly disperse.  This is accompanied by a continuous creation of
new arms. The gas density maxima, being the regions of strongest cloud
coagulation, can be identified with dust lanes accompanying spiral arms in
galaxies. Shock fronts, which are seen in hydrodynamical gas simulations
with a bar gravitational potential, could be identified with the gas density
enhancements seen in our simulations, despite the limited resolution of
our sticky particle model. The velocity in the corotating frame shows a very
similar behavior to the hydrodynamical models with a strong change in
velocity direction at the location where the shock is expected.

All calculated models yield qualitatively similar magnetic field
structures, so we decided to present figures mainly for the model A1. The
large-scale magnetic field quickly responds to non-axisymmetric velocity
perturbations of the gas flow in the bar and spiral arms. The magnetic field
is locally amplified by a local compression near leading sides of the bar and
inner edges of the spiral arms. It already forms magnetic arms at
early stages of evolution. As the gaseous arms dissipate, the magnetic
ones remain, often between newly formed gaseous arms.
In Fig.~\ref{bfeld} where magnetic vectors are superimposed onto the gas
density grey plot at the time step of $3.5 \times 10^8$~yr, we can see the
magnetic maxima related to newly created arms as well as those corresponding
to arm segments present some time ago. The magnetic arm amplified at the
earlier stages (about 100~Myr ago) is visible in the south-west part of
Fig.~\ref{bfeld}~(left). The pitch angles of the magnetic vectors
in the interarm space are similar to those of gaseous arms.
The time at which the results are prsented in the figures
was chosen so to provide the best resemblance to observations
shown in Fig.~\ref{n3627}.

In Fig.~\ref{bfeld}~(right) the velocity vectors (in a co-rotating frame)
are superimposed onto the magnetic intensity grey plot
showing the character of gaseous flow at the same time interval.
In region near the southern end of the bar we can see that 
the radial velocity changes its sign, causing the amplification
of the magnetic field there. Such regions are distributed along
gaseous non-axisymmetric features  at different time
steps and they could be identified with gaseous shocks.

\subsection{The polarization maps}

The map of polarization vectors for a simulated almost face-on ($30^\circ$)
galaxy (model A1, see Table \ref{tab2}), superimposed onto the contours of
polarized intensity obtained for the same time step ($3.5 \times 10^8$~yr)
as in Fig.~\ref{bfeld}, is presented in Fig.~\ref{fmaps35}. The resultant
distributions of polarized vectors and intensity look much smoother than
the vectors presented in Fig.~\ref{bfeld}, which is due to beam smoothing and
integration of magnetic field structures along the line of sight.
The most conspicuous features in these maps are the bright
polarized arms extending along the gaseous spirals and also in
the interarm regions. Fig.~\ref{fmaps35}~(left) presents a three-armed
structure of polarized ridges. Two of them coincide with the gas
density maxima constituting newly generated gaseous arms. Another one runs
from the SW bar end to the interarm space towards the east and traces an
older gaseous structure detached and dispersed 100~Myr ago. In this respect
the magnetic arm may trace old gaseous patterns which were generated and faded
away a long time ago in galactic history. In experiments with a high
diffusion coefficient some polarized emission also fills smoothly the interarm
space, as is often observed in spiral galaxies (Beck et al.~\cite{rb96}).

\begin{figure}
\resizebox{\hsize}{!}{\includegraphics{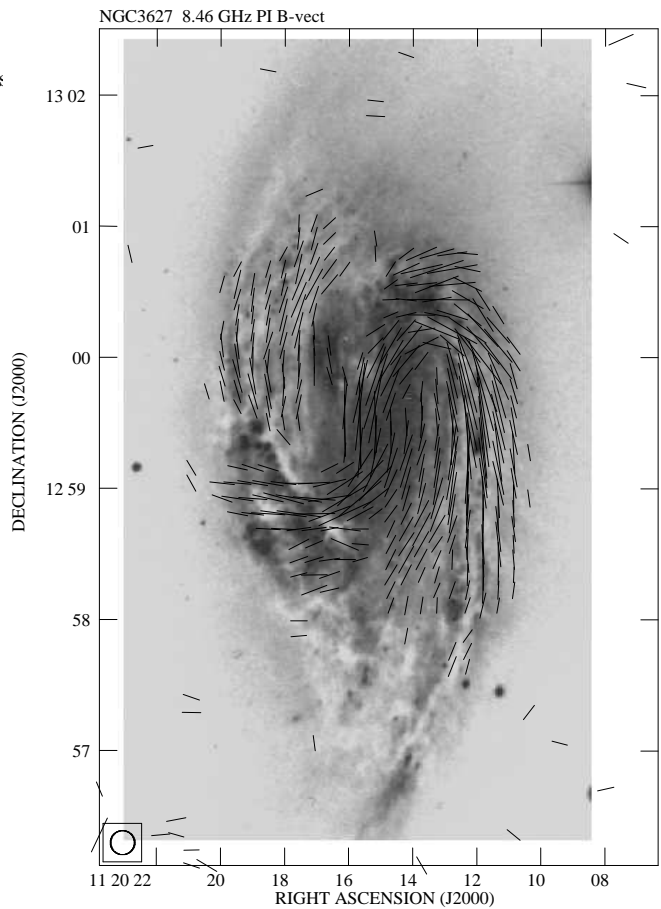}}
\caption{
The B-vectors of length proportional to polarized intensity of
NGC~3627 at 8.46~GHz superimposed onto the optical image by Arp
(\cite{arp66}). A vector length of 10\arcsec corresponds to a polarized
intensity of 50~$\mu$Jy/b.a. The orientations of vectors have
been corrected for Faraday rotation effects. The map resolution
is $11\arcsec$ (see Soida et al~\cite{soi01})
}
\label{n3627}
\end{figure}

The polarization B-vectors are generally aligned with gaseous spiral arms also
in the interarm space, forming a coherent spiral pattern with pitch angles
similar to what we identify as the dust lanes. However,
in the model A1 the B-vectors show locally strong pitch angle differences
between the arms and the interarm regions. This is particularly visible
for the region around R.A.$_{2000}:\rm 11^h 59^m 54^s$;
Dec$_{2000}: 14\degr 58\arcmin 30\arcsec$ where the B-vectors in the arm
and the interarm meet at almost right angles (Fig.~\ref{fmaps35}~(right)).
The interarm vectors run parallel to the bar turning suddenly towards the
orientation of a spiral arm, which resembles the situation in the barred
galaxy NGC~3627 as presented in Fig.~\ref{n3627} (Soida et al. \cite{soi01}).
We note particularly the region around R.A.$_{2000}\rm: 11^h 20^m 14^s$;
Dec$_{2000}: 13\degr 00\arcmin$ and that around
R.A.$_{2000}:\rm 11^h 20^m 18^s$; Dec$_{2000}: 12\degr 59\arcmin$ in
Fig.~\ref{n3627} where the arm and interarm B-vectors meet at right angles. 
We note also that in Fig.~\ref{fmaps35} around the
region of R.A.$_{2000}:\rm 11^h 59^m 55^s$;
Dec$_{2000}: 14\degr 59\arcmin 30\arcsec$, the B-vectors turn suddenly from 
the orientation {\it perpendicular} to the bar to that {\it parallel}
to the bar. This bears some (limited) resemblance to the situation in
NGC 1097 (Beck et al~\cite{bec99}). 
We note however that the latter galaxy shows a much stronger 
bar than that in our model. For this reason the validity of such a comparison
is still limited and will be a subject of a separate study involving
an analytically introduced bar potential.

\begin{figure}
\resizebox{\hsize}{!}{\includegraphics{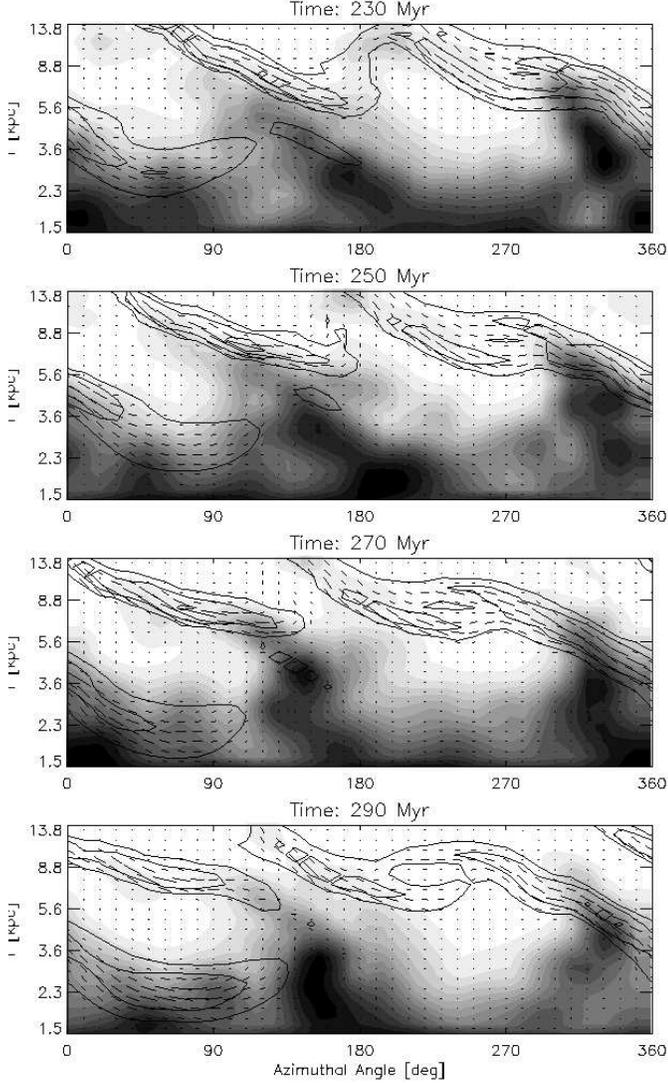}}
\caption{
Contour plots of polarized intensity maps and B-vectors from model B
superimposed onto the gas density maps (integrated along the line of sight)
shown in the azimuth -- $\log (r)$ frame at four time steps. 
}
\label{urbs}
\end{figure}

The bright polarized arms can be conveniently analyzed in the frame of
azimuthal angle in the disk with respect to the bar orientation and
$\ln(r)$, in which a logarithmic spiral appears as a straight line
inclined by its pitch angle.  We analyze both the position of the
arms relative to the bar and the orientations of polarization B-vectors.
Fig.~\ref{urbs} shows the time evolution of the model B at four time
steps:  2.3, 2.5, 2.7 and $2.9 \times 10^8$~yr. A greyplot
of the gas density (logarithmic scale) is included, as well. In our frame the
gaseous structures in the inner disk (thus more directly connected to the bar)
do not change their position much. However, weak gas density maxima in the
outer disk ($r>8$ kpc) systematically drift to the left in Fig.~\ref{urbs}, 
being thus delayed with respect to the bar rotation. The inner parts
of the magnetic arms remain tied to the gaseous structures and do not
change the position much with respect to the bar.
An obvious drift of magnetic structures occurs in the outer disk
(Fig.~\ref{urbs}).  A kind of ``magnetic wave'' move to the left, being
continuously generated at the bar ends.

\begin{figure}
\resizebox{\hsize}{!}{\includegraphics{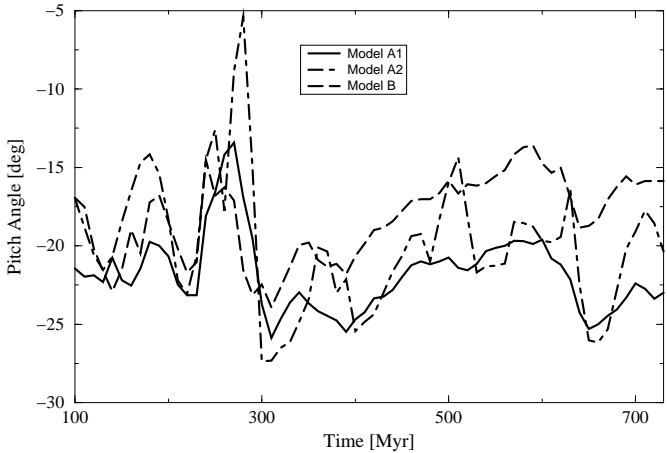}}
\caption{
The evolution of mean magnetic pitch angle for models A1, A2 and B,
averaged in galactocentric rings between 8 and 12~kpc.
}
\label{pitch}
\end{figure}  

In order to compare the time evolution of pitch angles of polarization 
B-vectors obtained for all models, we averaged the pitch angle values between
the radius $r=8$~kpc and 12~kpc along the azimuthal direction for all
time steps (Fig.~\ref{pitch}). For all models the absolute value of
the pitch angle is quite high -- from about $18\degr\pm 2.6\degr$
for model B (without the dynamo mechanism) to about $22\degr\pm 2.5\degr$
for our base model (A1).  For the simulations with smaller dynamo
coefficients after 300~Myr (model A2) the mean pitch angle grows with time
from $-17\degr$ to about $-22\degr$ with a std. dev. of $\pm 4\degr$.

\begin{figure}
\resizebox{\hsize}{!}{\includegraphics{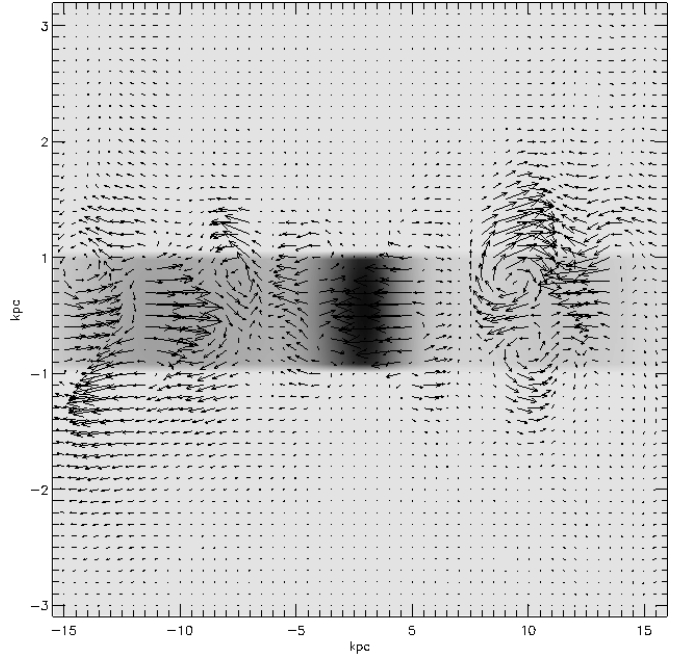}}
\caption{
The magnetic field vectors and gas density shown in the plane 
perpendicular to the galactic plane at time step 490~Myr. 
}
\label{bxz}
\end{figure}

Fig.~\ref{bxz} presents the structure of the poloidal magnetic field at the
disk/halo interface for the evolution time of 490~Myr for model A1 in the
plane perpendicular to the galactic plane. Magnetic structures in the halo
are formed quite quickly and they persist over most of the evolution time.
The excited mode is mainly of an S0 type, which is a consequence of the
chosen form of the dynamo tensor $\alpha$. In Fig.~\ref{bxz} we can see that
the magnetic lines of force form separate loops under and above the
galactic plane, having opposite senses of circulation. Instead of single,
galaxy-scale loop-like structures we obtained rather smaller, localized 
structures centered at various heights up to 1.5~kpc above the disk plane.
The growing value of the magnetic diffusion in the halo does not allow 
the formation of stronger magnetic structures higher than 2~kpc above
the disk plane. The calculations with the decreasing rotational velocity
(model C, see Table 2) in the halo give no significant differences between
models having the same value of the other dynamo parameters
(Brandenburg et al. 1993).

\begin{figure}
\resizebox{\hsize}{!}{\includegraphics{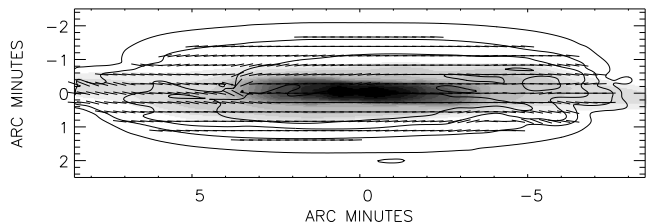}}
\caption{
The edge-on (inclination of 85$\degr$) polarized intensity map with
B-vector map from model A1 at time 510~Myr. 1\arcmin{} corresponds to
about 2~kpc in the modeled galaxy
}
\label{emaps}
\end{figure}

Fig.~\ref{emaps} presents the edge-on view of resultant polarization B-vectors
overlaid onto the polarization intensity contours projected almost edge-on
($\rm inclination=85\degr$) onto the sky for the model A1 at the time step
510~Myr. The polarization map shows the vectors which are parallel to
the galactic disk. This is caused by the fact that we see only the parts
of poloidal loops closest to the plane where there are enough cosmic ray
electrons to generate significant synchrotron emission. The projection and
beam smoothing effects efficiently blur all localized vertical fields
leaving only its mean plane-parallel component seen in polarization.
This resembles the majority of galaxies (except NGC~4631,
Golla \& Hummel \cite{gol94}), in which a disk-parallel magnetic field is
observed close to the disk plane (Dumke et al. \cite{dum95},
Dumke et al. \cite{dum00}, Beck et al.~\cite{rb96})

\begin{figure}
\resizebox{\hsize}{!}{\includegraphics{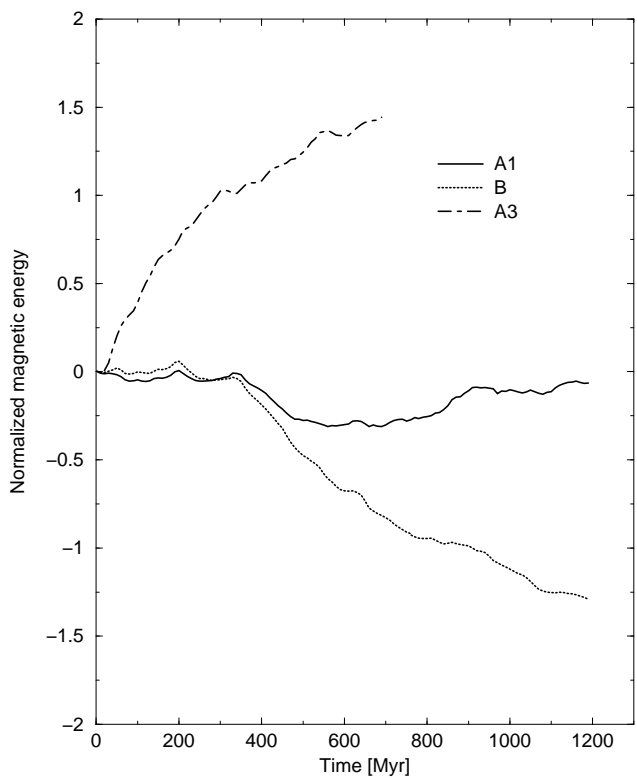}}
\caption{
Time evolution plots of the total magnetic energy normalized
to the initial value (in logarithmic scale)
for models A1, A3 and B. The initial field was the saturated field calculated
with the primary velocity for the parameters of model A1.
}
\label{btot}
\end{figure}

Fig.~\ref{btot} shows the magnetic energy evolution for the models
A1, A3 and B. We started our simulations with an already evolved dynamo
magnetic field rather than with a weak seed field, therefore the energy
of the model A1 remains nearly constant. The magnetic energy of model B
with the same diffusivity decreases, because no further turbulent dynamo
action occurs. Surprisingly, the magnetic energy of model A3 with a very
small value of $\alpha$ and with reduced turbulent diffusivity is
increasing. The dynamo number has the same value as in model A1. Therefore
the magnetic energy should remain constant, similar to model A1. The only
explanation for the growth is an amplification by the velocity. At that
point it is difficult to decide if the velocity itself is acting as a dynamo.
Because of a nearly linear growth in the magnetic field strength (square root
of the energy) it could be a temporary increase of magnetic energy due to
the shear flow. The vertical diffusion time for that model is about 10~Gyr. 
A consequence would be growth of the magnetic field without a major
influence of $\alpha$. A quenched $\alpha$ and $\eta$ would give
a further increase of magnetic energy, due to complicated 3D velocity 
flows. The surprise is that the pitch angle remains large even
with a weak influence of $\alpha$. Further investigation of the magnetic
field evolution in a lower diffusivity medium will be presented in a 
forthcoming paper. 

\section{Discussion}

Our experiments clearly indicate that a strong magnetic component
associated with spiral arms persists in spite of introducing the large gaseous
halo and of the fact that spiral arm perturbations occur only in a thin disk.

The most interesting result is the fact that during the whole evolution time
the modeled polarized intensity forms wave-like maxima apparently created
in the compresion regions in the disk, seen as the gaseous arms.
These gaseous spirals do not corotate with the bar but have a somewhat
smaller angular velocity. This result is in agreement with earlier numerical
works that proposed spiral arms  having a pattern speed smaller than that
of a bar (e.g. Sellwood \& Sparke \cite{sel88}, Lindblad et al. \cite{lin96},
Sellwood \& Wilkinson \cite{sel93}, Tagger et al. \cite{tag87}).

The most astonishing result is that the magnetic spiral pattern rotates
slightly slower than the gaseous spirals almost independent of the diffusivity. 
This is a completely new result. This process causes a drift of magnetic
field maxima from the gaseous spirals to the interarm space. This bears some
quantitative resemblance to e.g. NGC~6946 (a weakly barred galaxy,
Beck \& Hoernes~\cite{bec96}). Due to this mechanism magnetic arms could
work as a tracer of spiral arms present at earlier time stages. For the first
time we obtained rapid turning of the magnetic pitch angle between arm and
interarm regions, which is also observed in NGC~3627 (Soida et
al. \cite{soi01}) and in NGC~1365 (Shoutenkov et al. 2000, 
Beck et al. in prep, Lindblad 1999). This gives the impression of two
separate components, one in spiral arms and one between them, as suggested
by some observers (Soida et al. \cite{soi01}). Simulations with different 
halo-to-disk mass ratios, resulting in a spiral pattern without a bar,
did not show any difference between magnetic and optical pattern speeds
(Elstner et al. \cite{els00}). 

In contrast to our previous works the resultant magnetic field structure is
fully three-dimensional with poloidal loops extending into the halo. Their
existence does not contradict observations showing disk-parallel polarization
vectors close to the disk plane (Dumke et al. \cite{dum95}, Dumke et
al. \cite{dum00}). We note however a limited applicability of our model
to particular real galaxies as we did not introduce concrete values
of parameters to reproduce particular observations. We restrict ourselves
to mentioning phenomenological features which may be worth further detailed
studies to explain the observations in more detail.

\section{Summary and conclusions}

The evolution of large-scale galactic magnetic fields in a galaxy with a
gaseous halo and a bar has been demonstrated using 3D numerical simulations.
In order to solve the dynamo equation we used velocities obtained from a
self-consistent N-body, sticky particle code. 
We found that: 

\begin{itemize}
\item[1.]{The magnetic arms  persist in the presence of a large
halo enabling a large-scale three-dimensional field.}
\item[2.]{The magnetic arms are also present in interarm regions, in
agreement with observations.}
\item[3.]{Our calculations confirm the fact, known from 
earlier papers, that the pattern speed of the bar is
higher than the pattern speed of spiral arms.}
\item[4.]{The magnetic field may take the form of a
``magnetic wave'' with magnetic arms drifting into the interarm
areas due to dynamical dissipation and shear amplification 
independent of the value of the turbulent diffusion.} 
\item[5.]{The magnetic pitch angles for all models keep the mean value between  
$-15\degr$ and $-25\degr$ independent of the diffusivity.}
\item[6.]{The edge-on view of the modeled magnetic field agrees with
the majority of real observations.}
\end{itemize}

\acknowledgements{The authors express their gratitude to Dr. Susanne von Linden
for kindly supplying the velocity fields and to Dr. Anvar Shukurov for his
valuable comments. This work was partly supported by a
grant from the Polish Committee for Scientific Research (KBN), grant no.
4264/P03/99/17 and computing grant KBN/SPP/UJ/011/1996 }


\begin{thebibliography}{}
\bibitem[1966]{arp66} Arp H.: 1966, ApJS 14, 1
\bibitem[1996]{bec96}
Beck R., Hoernes P.: 1996, Nature 379, 47
\bibitem[1996]{rb96}
Beck R., Brandenburg A., Moss D., Shukurov A.M., Sokoloff D.D.: 1996,
        Ann. Rev. A\&A 34, 155
\bibitem[1999]{bec99} Beck R., Ehle M., Shoutenkov V., Shukurov
  A., Sokoloff D.D.: 1999, Nat. 397, 324
\bibitem[1993]{bran93}
Brandenburg A., Donner K. J., Moss D., Shukurov A.,            
Sokoloff D. D., Tuominen I.: 1993,  A\&A 271, 36 
\bibitem[1985]{com85}
Combes F., G\'erin M.: 1985, A\&A 150, 327
\bibitem[1995]{dum95}
Dumke M., Krause M., Wielebinski R., Klein U.: 1995, A\&A
	    302, 691
\bibitem[2000]{dum00}
Dumke M., Krause M., Wielebinski R.: 2000, A\&A 355, 512
\bibitem[1998]{els98}
Elstner D., Lesch H., von Linden S., Otmianowska-Mazur K.,
 Urbanik M.: 1998, Studia geoph. et geod. 42, 373
\bibitem[2000]{els00}
Elstner D., Otmianowska-Mazur K., v. Linden S., 
Urbanik M.: 2000, A\&A 357, 129
\bibitem[1997]{engl97}
Englmaier P., Gerhard O.: 1997, MNRAS 287, 57
\bibitem[1996]{fro96}
Fr\"ohlich H.-E., Schultz M.: 1996, A\&A 311, 451
\bibitem[1994]{gol94}
Golla G. and Hummel K.: 1994, A\&A 284, 777
\bibitem[1998]{kal98}
Kalberla P.M.W., Kerp J.: 1998, A\&A 339, 745.
\bibitem[1996]{lin96}
Lindblad P.A.B., Lindblad P.O., Athanassoula E.: 1996, A\&A 313, 65
\bibitem[1996]{lin99}
Lindblad P.A.B.: 1999, A\&A Rev. 9, 221
\bibitem[1998]{lin98}
von Linden S., Otmianowska-Mazur K., Lesch H., Skupniewicz G.: 1998, 
 A\&A 333, 79
\bibitem[1997]{les97}
Lesch H. and Chiba M.: 1997, Fundamentals of Cosmic Physics 18, 273
\bibitem[1998]{mos98}
Moss D.: 1998, MNRAS 297, 860
\bibitem[1998]{mosetal98}
Moss D., Korpi M., Rautiainen P., Salo H.: 1998, A\&A 329, 895 
\bibitem[1999]{mos99}
Moss D., Rautiainen P., Salo H.: 1999, MNRAS 303, 125
\bibitem[1997]{otm97}
Otmianowska-Mazur K., von Linden S., Lesch H., Skupniewicz G.:
 1997, A\&A 323, 56 
\bibitem[2000]{otm00}
Otmianowska-Mazur K., Chy\.zy K., Soida M.: 2000, A\&A 359, 29
\bibitem[1970]{pac70}
Pacholczyk A.G.: 1970, ``Radio Astrophysics. Nonthermal Processes in
 Galactic and Extragalactic Sources.'', W.H. Freeman Co., San Francisco. 
\bibitem[1998]{roh98}
Rohde R., Elstner D.: 1998, A\&A 333, 27
\bibitem[1999]{roh99}
Rohde R., Beck R., Elstner D.: 1999, A\&A 350, 423 
\bibitem[1988]{sel88} 
Sellwood J.A., Sparke L.S.: 1988, MNRAS 231, 25
\bibitem[1993]{sel93}
Sellwood J.A., Wilkinson A.: 1993, Rep. Prog. Phys. 56, 173
\bibitem[2001]{soi01}
Soida M., Urbanik M., Beck R., Wielebinski R., Balkowski C.: 2001, A\&A 378, 40
\bibitem[2000]{}
Shoutenkov V., Beck R., Shukurov A., Sokoloff, D.: 2000,
      in "The Origins of Galactic Magnetic Fields", 24th meeting
      of the IAU, Joint Discussion 14, August 2000, Manchester, England,
      p.34
\bibitem[1992]{sto92}
Stone J.M., Norman M.L.: 1992, ApJS 80, 791
\bibitem[1987]{tag87}
Tagger M., Sygnet J.F., Athanassoula E., Pellat R.: 1987,
ApJ 318, L43
\bibitem[2000]{tul00}
T\"ullmann R., Dettmar R.-J., Soida M., Urbanik M., Rossa J.: 2000,
A\&A 364, L36-L41

\end{thebibliography}
\end{document}